\documentstyle[pra,aps,amsmath,amssymb,psfrag,graphicx,epsfig,changebar,psfrag]{revtex}

\def\>{\rangle}
\def\<{\langle}
\def\n{\nonumber}
\def\lk{\left\{}
\def\rk{\right\}}
\def\<{\langle}
\def\>{\rangle}
\def\be#1\ee{\begin{equation}#1\end{equation}}
\def\ba{\begin{eqnarray}}
\def\ea{\end{eqnarray}}

\input{psfig}

\begin{document}
\twocolumn
\widetext

\title{Failure of Effective Potential Approach: Nucleus-Electron Entanglement in the He$^+$-Ion.}
\author{J. Gemmer and G. Mahler}
\address{Institut f\"ur Theoretische Physik,\\Universit\"at Stuttgart, Pfaffenwaldring 57,\\70550 Stuttgart, Germany}
\date{\today}
\maketitle


\begin{abstract}

Entanglement may be considered a resource  for quantum-information
processing, as the origin of robust and universal equilibrium
behaviour, but also as a limit to the validity of an effective
potential approach, in which the influence of certain interacting
subsystems is treated as a potential. Here we show that a closed three particle (two protons, one electron)
model of a He$^+$-ion featuring realistic size, interactions and energy scales of
electron and nucleus, respectively, exhibits
different types of dynamics depending on the initial state: For some
cases the traditional approach, in which the nucleus only
appears as the center of a Coulomb potential, is valid, in others this
approach fails due to entanglement
arising on a short time-scale. Eventually the system can even show
signatures of thermodynamical behaviour, i.e. the electron may relax
to a maximum local entropy state which is, to some extent, independent of
the details of the initial state.

\end{abstract}

\pacs{PACS number(s): 03.65.-w, 05.30.-d, 05.70.Ln}

\narrowtext
During the last decades entanglement has attracted considerable interest in a lot of fields of quantum mechanical research. In the
begining, after the publication of the famous EPR-paradox \cite{EIN35}, it
was mostly considered a philosophical puzzle, challenging the basic
principles of quantum physics itself.\\
Later on, methods to deliberately produce and detect entanglement were
developed, originally in the field of quantum optics \cite{CHU98,TUR98}.\\
After that, the idea of exploiting entanglement in a technical sense
became popular and a number of so-called quantum algorithms have been
suggested, that could outperform corresponding classical algorithms \cite{DEU85,GRO97,SHO94}. Those algorithms require the controlled manipulation
of many entangled subsystems. Although there is presently much research directed towards those ends, no large scale
quantum computers are in sight so far; obviously, the engineering of entanglement is a
very hard problem \cite{ALI01}.\\
Probably because entanglement is so difficult to control and its
consequences may seem puzzling, it has rarely been discussed in
the context of ``effective potentials''. However,
taking a closer look
one has to admit that there are very few true ``single system''
scenarios. Even the historical double slit or the famous
``particle-in-a-box'' problem are, in fact, compound systems, if one starts from
first principles, for the wall with the slits or the box are sytems
consisting of many particles themselves, that should be described by
wavefunctions. Almost all potentials (even so-called ``classical''
ones, underlying, e.g., mesoscopic or microscopic structure models) are due to interacting subsystems. So, the question arises, why the standard effective
potential approach seems to be succesful in so many cases, despite
entanglement being
fairly generic, whenever systems interact \cite{BRA01,GEM01A}.\\
On the other hand, this generic nature, together with the fact that entanglement leads to increased
entropy for the entangled  subsystems, has even led to the idea that
entanglement with the surrounding could be responsible also for the
validity of the second law for thermodynamical systems \cite{GEM01}. It is, therefore, tempting to look for possible transitions between
thermodynamic and microscopic behaviour, i. e. between macro- and micro-control\cite{SCI01}.\\
For this purpose we are going to study a closed bi-partite quantum
system hierarchically grouped into a tightly bound pair, which
approximately generates an effective potential for the third, weakly
bound particle. Thus we want to present here a concrete example
for a system showing entanglement for certain initial states, after the
classical effective potential has been replaced by a subsystem with
internal degrees of freedom. This example is the He$^+$-ion and the subsystem the nucleus itself. This object has been
chosen for two reasons: Firstly, one can come up with a fairly simple model, which does
not require complicated numerical analysis, and is nevertheless
reasonably close to reality. Secondly, the nuclear and electronic
excitations represent so different energy scales that entanglement
appears to be beyond reasonable expectation. While this is, indeed,
correct from an experimental point of view, it may, nevertheless, come as a surprise that mathematically ``typical'' initial states will,
indeed, lead to entanglement.\\
The He$^+$-ion consists, on the level of nucleons and electrons, of five
particles. Since the neutrons do not feel the Coulomb force they are neglected here. The remaining three particles, which are
relevant for our model are described by the following Jacobi-coordinates as
sketched in {\bf Fig.1}:
\ba
\vec{x}_e&=&\vec{S}+\frac{2M}{2M+m}\vec{R}\\
\vec{x}_{p_1}&=&\vec{S}-\frac{m}{2M+m}\vec{R}+\frac{1}{2}\vec{K}\n\\
\vec{x}_{p_2}&=&\vec{S}-\frac{m}{2M+m}\vec{R}-\frac{1}{2}\vec{K}.\n
\ea
Here $M$ is the proton-mass and $m$ the electron-mass.\\
These coordinates are chosen to decouple the
center of mass degree of freedom from the others. The Hamiltonian reads:
\ba
\label{ham}
&&\hat{H}=-\frac{\hbar^2}{2(2M+m)}\nabla^2_{\vec{S}}-\frac{\hbar^2}{M}\nabla^2_{\vec{K}}+V(\vec{K})-\\
&&-\frac{\hbar^2(2M+m)}{4Mm}\nabla^2_{\vec{R}}-\alpha\hbar c\left(\frac{1}{\left|\vec{R}-\frac{1}{2}\vec{K}\right|}+\frac{1}{\left|\vec{R}+\frac{1}{2}\vec{K}\right|}\right)\n
\ea
where $\alpha$ is the fine structure constant and  $V(\vec{K})$
describing the internuclear force is chosen to be:
\be
V(\vec{K}):= \left\{ \begin{array}{ccc}0&:&|\vec{K}|\leq 2r_0\\
\infty&:&|\vec{K}|> 2r_0 \end{array} \right.,
\ee
with $r_0\approx2,2\cdot10^{-15}m$ \cite{MAY85} taken as the radius
of $\alpha$-particles obtained from scattering experiments.\\
After the center of mass degree of freedom has been separated, there
is, obviously, no way of decoupling the Hamiltonian in terms of $\vec{K}$
and $\vec{R}$ by means of any further transformation (non-separability).\\
In a classical analysis one would always argue that $|\vec{K}| \ll
|\vec{R}|$ and therefore expand the last part of the Hamiltonian
neglecting all higher order terms and thus decouple the
Hamiltonian completely. We will eventually do something similar, but boldly
applying the same argument in quantum mechanics would miss the crucial
point, as will be seen below.\\
The idea now is to analyse this model using perturbation
theory. Therefore we have to divide the Hamiltonian into a main
($\hat{H}_0$) and a perturbative ($\hat{H}_1$)
part. Simply taking the last term in (\ref{ham}) as the perturbation
would defenitely not be good enough, for the electron would be free
according to $\hat{H}_0$, and one could not expect the perturbation
series to converge. It seems more promising to introduce an effective
potential, which models the mean effect of the nucleus on the electron
in $\hat{H}_0$, and take the deviation of the ``real'' interaction from
this effective potential as the perturbation. Following these ideas
and putting aside the center of mass degree of freedom, we get:
\ba
\hat{H}_0&=&-\frac{\hbar^2}{M}\nabla^2_{\vec{K}}+V(\vec{K})-\frac{\hbar^2(2M+m)}{4Mm}\nabla^2_{\vec{R}}-\frac{2\alpha\hbar c}{\left|\vec{R}\right|}\n\\
\hat{H}_1&=&-\alpha\hbar c\left(\frac{1}{\left|\vec{R}-\frac{1}{2}\vec{K}\right|}+\frac{1}{\left|\vec{R}+\frac{1}{2}\vec{K}\right|}-\frac{2}{\left|\vec{R}\right|}\right)
\ea
Obviously, the full Hamiltonian has not changed, but
$\hat{H}_0$ describes now two decoupled systems, the nucleus, $c$, alone and
the electron, $e$, like it is usually treated, bound by a potential that
would arise if the nucleus was a pointlike object without any
substructure. Taking into account  that the ratio of the radius of the
nucleus to a typical distance of the electron from the nucleus (say,
the Bohr radius) is smaller than $10^{-4}$ \cite{MAY85,BRA85} one would expect that
$\hat{H}_0$ should already give a pretty good picture of the real
system, so that the effect of $\hat{H}_1$ should be comparatively
small. To verify
this, one needs to calculate the perturbation matrix. This analysis,
which is straightforward, but too long to be presented here,
shows, that all its entries are much smaller than the corresponding
energy differences of the electron system (roughly by a factor of
$10^{-10}$) and that it is almost diagonal within the degenerate
eigenspaces of $\hat{H}_0$. This means that the
(product)energy eigenstates of the unperturbed problem remain
pratically unchanged, what matters are the corrections to the energy eigenvalues, induced by the effective coupling.\\
For states with vanishing orbital angular momentum of the electron
system as well as of the nucleus system, those corrections can be
calculated analytically to first order:
\be
\label{corr}
E^1_{nN}=\frac{\alpha\hbar cr_0^2}{6(an)^3(n!)^2}\left(\frac{1}{3}-\frac{1}{2\pi ^2N^2}\right)
\ee
Here, $a$ is the Bohr radius, $n$ is the principal quantum number of
the electron, $N$ that of the
nucleus. These corrections are at most on the order of $10^{-10}$eV which is extremly
small compared to the energy-scale of the uncoupled
system. Nevertheless, they are nonadditive (which could not happen if
the perturbation was local) and can, though being very small, cause
entanglement.\\
Entanglement measures for general multi-partite systems are still
under dispute \cite{VED97}. However, if the state of the whole system is a pure state, and the full
system is being regarded as divided into two subsystems, a convenient
entanglement measure is $1-P_e$, with
\be
\label{pur}
P_e(t)=\mbox{Tr}\lk \hat{\rho}^2_e(t)\rk =\mbox{Tr}_e\lk \left(\mbox{Tr}_c\lk|\psi(t)\>\< \psi(t)|\rk \right)^2\rk,
\ee
where $\hat{\rho}_e$ is the reduced density operator of subsystem $e$
and $|\psi(t)\>$ is the wave function of the total system. Under these
conditions this quantity yields the same value, no matter for  which
subsystem it is calculated, i. e., $P_e=P_c=:P$. $P$ is
called the purity since it takes on its maximum value $1$ if the
subsystem is in a pure state. Furthermore, $P$ can be used as a
criterion for a subsystem to act as an effective potential for the
other subsystem: $P$ would have to be $1$ for all times. The more $P$
deviates from $1$ the larger is the error that would occur if a
Hartree-type description was used to calculate the evolution.\\ 
The {\bf Figs. 2-3} display the time evolution of the purity $P(t)$ for
some specific initial product states of vanishing orbital angular
momentum. In the following the numbers in the ``ket'' symbols give the
principal quantum numbers of the electron ($n$) and the nucleus
($N$).\\
The purity-evolution we find for the initial state
$|\psi(0)\>=\frac{1}{2}(|1\>+|2\>)^c\otimes(|1\>+|2\>)^e$ is displayed
in {\bf Fig.2}. Obviously, considerable entanglement is being built up on the
$10^{-5}$ second timescale! The fact that the mean distance between the
particles making up the nucleus is much smaller than the mean
distance of the electron from the nucleus does not mean that the
nucleus may necessarily be treated as an effective potential. What
really matters is whether or not the electron ``feels'' the different
potentials the nucleus creates in its different energy eigenstates. And,
obviously, already a slight difference destroys local coherence
quite rapidly. Since the energy-transfer from the electron to the
nucleus (or vice versa) can be neglected in this model, the dynamics
are effectively restricted to the space of the four states occupied in the
initial state. Within this effective two-level-two-system-subspace a maximum
entangled state, an EPR-state, is implemented at the
minima of the purity.\\ 
In the initial state
$|\psi(0)\>=\frac{1}{2}(|1\>+|2\>)^c\otimes(|14\>+|15\>)^e$ the electron
system is now in a superposition of states with higher principal
quantum numbers (14,15 instead of 1,2) which reduces the probability
to find the electron in the direct vicinity of the nucleus
drastically. This difference shows in the purity-evolution: In
principle it looks the same as the one in {\bf Fig.2}, but instead of
a timescale of $10^{-5}$ s, the purity reaches now values that
differ signifficantly from $P=1$ on a timescale of  $10^{19}$ s (about the age of the universe!). Hence for all practical purposes an
effective potential approach will yield excellent results in this
case.\\
Yet a different situation arises if the initial state consists of
superpositions  of somewhat more than two energy eigenstates. {\bf Fig.3} shows the purity-evolution of the initial state
$|\psi(0)\>=\frac{1}{10}(\sum_{N=1}^{10}|N\>)^c\otimes(\sum_{n=1}^{10}|n\>)^e$.
In this case, the purity no longer oscillates but decays on an
intermediate time-scale of $1$ sec. to reach a final value of $\bar{P} \approx
0.19$. Note that this behaviour occurs even though the dynamics of the
whole system is perfectly unitary. It is also ``universal'' in the
sense that it appears independently of the phases of the amplitudes of
the initial state. Since small purity values
correspond to large von Neumann entropies, this behaviour is very much like the behaviour of a
true thermodynamical system, meaning the relaxation of the system into
its maximum entropy state.\\
But to obtain this $\bar{P}$ one would not even have to analize the full Schr\"odinger dynamics, for the equilibrium value of the purity can be calculated from the following formula \cite{GEM01}
\be
\label{for}
P_{eq}=\sum_n(W_n^e)^2+\sum_N(W_N^c)^2-\left(\sum_n(W_n^e)^2\right)\left(\sum_N(W_N^c)^2\right)
\ee
Here $W_n^e$ and $W_N^c$ denote the initial-state-probabilities of finding
the electron in state $n$, the proton subsystem in state $N$. For the
case of {\bf Fig.3} we have $W_n^e=W_N^c=\frac{1}{10},(n,N=1,2...10)$
and thus $P_{eq}=0.19$ in surprisingly good agreement with the above numerical result. The
derivation of eq.(\ref{for}) does not contain any dynamical details, but is merely based on unbiased averages of the respective compound
Hilbert space and expected to be valid for certain classes of
thermodynamical systems \cite{GEM01B}.\\
Classical degrees of freedom are usually associated with the number of
independent coordinates, which is proportional to the dimension of
classical phase-space. Even if we take as a quantum analogue the
dimension, $d$, of the respective Hilbert-space, our present model
with the specific initial states considered is still far from the
thermodynamical limit ($d\gg 1$). Neverthelss, the model clearly shows signatures of
thermodynamical (statistical) behaviour for a large class of initial
states. These features would become more and more significant as more and
more energy eigenstates were superimposed in the initial state.\\
Thus, as already found by R. Jensen and R. Shankar \cite{JEN85}, it is not necessary for a system to have many classical degrees of freedom in order to exhibit statistical
behaviour. What
does seem to be necessary is the coupling of the system to another
system, such that a sizable part of the Hilbert space of the coupled
system is accessible, even though the coupling might be weak and
without energy-exchange.\\
A class of initial states for which practically no entanglement
will ever arise is the set of product states without any
superposition of energy-eigenstates in the nucleus. Although this class
might not be large from a theoretical point of view, it contains the
most common states realizable in the laboratory, namely the case of the nucleus
being in the ground state. Since excitations of the nucleus are usually
in the $MeV$ regime, the nucleus will decay into the ground state due
to the coupling to the electromagnetic field, on a timescale much shorter than the one on which
entanglement arises. This coupling is completely absent in our present
model. It would thus be extremely difficult to
detect the predicted behaviour directly, even if one was able to put the
nucleus into the required superposition.\\
Alternatively, one might look for spectroscopic features resulting
from the energy-corrections given by eq.(\ref{corr}). To enhance the
effect under consideration larger nuclei or muonic atoms might be of
help. Of course, other three-particle models could
be selected for much easier experimental access \cite{HAN95} but also for less
``surprise''. Basically, though, our model represents a
Gedanken-experiment, designed to show that entanglement can indeed
appear ``almost everywhere''.\\
In conclusion, we have shown, based on a hierarchical bi-partite quantum
network, that there is a time period for which, starting from locally
pure states, the effective potential approach works for all practical
purposes. The length of this period depends sensitively on the initial
state and might very well approach infinity for typical states
accessible to experimentation. But there are also initial states for
which this period is short enough for the entanglement to built up on a relevant timescale. For some initial
states this entanglement will even lead to locally irreversible
equillibrium behaviour, controlled by thermodynamical laws. Thus the
standard Schr\"odinger equation can lead to both, unitary microscopic
and thermodynamic behaviour, depending on the experimental setup.\\
Our findings thus question a rather common view for modelling quantum systems: That there are
effective potentials for (quantum-) control and, independently, a bath to
account for decoherence. Though the latter may, indeed, be present,
typically, those two functions cannot be separated, control itself already implies
``de-control'': If even a nucleus can cause decoherence with respect to
an electron, then hardly anything needs to be save in this respect! 
We thank A.~Otte, I.~Kim, F.~Tonner M.~Stollsteimer, T.~Wahl,
T.~Haury, M.~Michel, P.~Borowski and H.~Schmidt for fruitful
discussions. Financial support by the Deutsche Forschungsgemeinschaft
is gratefully acknowledged.

\begin{figure}
\begin{center}
\psfrag{P1}{\hspace*{-3mm}proton 1}
\psfrag{P2}{\hspace*{-3mm}proton 2}
\psfrag{E}{electron}
\psfrag{R}{$\vec{R}$}
\psfrag{K}{$\vec{K}$}
\psfrag{S}{$\hspace{-2mm}\vec{S}$}
\psfrag{x1}{\raisebox{-5mm}{$\vec{x}_{p_1}$}}
\psfrag{x2}{$\hspace{-5mm}\vec{x}_{p_2}$}
\psfrag{xe}{$\hspace{2mm}\vec{x}_e$}
\psfrag{x}{$y$}
\psfrag{y}{$x$}
\psfrag{z}{$\hspace{-1mm}z$}
\epsfig{file=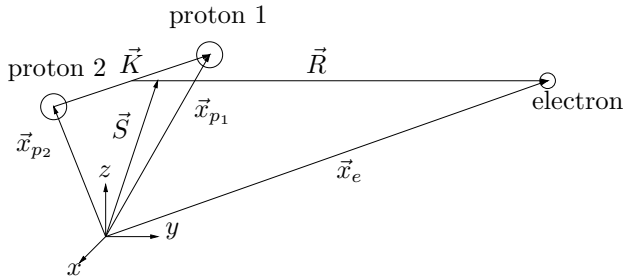, width=7cm}
\caption{Coordinates of the model: $\vec{S}$ denotes the
center-of-mass of the total system . Note: $|\vec{K}|:|\vec{R}|\approx 10^{-4}$}
\end{center}
\end{figure}
       
\begin{figure}
\begin{center}
\psfrag{x}{$t/10^{-5}$s}
\psfrag{y}{$P(t)$}
\psfrag{0.00001}{1}
\psfrag{0.00002}{2}
\psfrag{0.00003}{3}
\psfrag{0.00004}{4}
\psfrag{0.00005}{5}
\epsfig{file=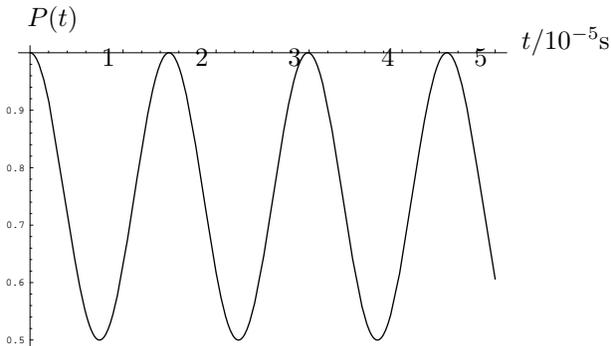, width=7cm}
\caption{Oscillating purity. Initial product state containing the two
lowest subsystem-states each, see text.}
\end{center}
\end{figure}

\begin{figure}
\begin{center}
\psfrag{x}{t/s}
\psfrag{y}{$P(t)$}
\epsfig{file=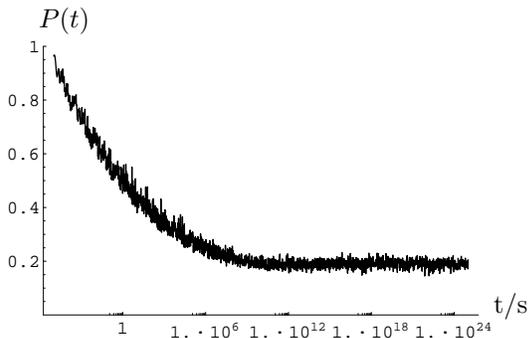, width=7cm}
\caption{Decaying purity. Initial product state containing the ten
lowest subsystem-states each, see text.}
\end{center}
\end{figure}

\end{document}